\documentclass[11pt]{article}
\title{Partial-indistinguishability obfuscation using braids}

\usepackage{graphicx, amssymb, amsmath, amsthm, fullpage, dsfont, hyperref}

\author{Gorjan Alagic\footnote{Institute for Quantum Information,
    California Institute of Technology, Pasadena, CA, USA. \texttt{galagic@gmail.com}}, \ Stacey
  Jeffery\footnote{Institute for Quantum Computing, University of
    Waterloo, Waterloo, ON, Canada. \texttt{smjeffery@gmail.com}}, \ and Stephen P. Jordan\footnote{National
    Institute of Standards and Technology, Gaithersburg, MD, USA. \texttt{stephen.jordan@nist.gov}}}

\date{}

\input{Qcircuit}

\newcommand{\circu}{\mathsf{C}}
\newcommand{\braid}{\mathsf{B}}
\newcommand{\normf}{\mathsf{N}}

\begin{document}
\bibliographystyle{plain}
\maketitle

\newcommand{\id}{\mathds{1}}
\newtheorem{lemma}{Lemma}
\newtheorem{corollary}{Corollary}
\newtheorem{proposition}{Proposition}
\newtheorem{theorem}{Theorem}
\newtheorem{definition}{Definition}
\newtheorem{conjecture}{Conjecture}
\newtheorem{algorithm}{Algorithm}
\newtheorem{protocol}{Protocol}

\begin{abstract}
An obfuscator is an algorithm that translates circuits into functionally-equivalent similarly-sized circuits that are hard to understand. Efficient obfuscators would have many applications in cryptography. Until recently, theoretical progress has mainly been limited to no-go results. Recent works have proposed the first efficient obfuscation algorithms for classical logic circuits, based on a notion of indistinguishability against polynomial-time adversaries. In this work, we propose a new notion of obfuscation, which we call partial-indistinguishability. This notion is based on computationally universal groups with efficiently computable normal forms, and appears to be incomparable with existing definitions. We describe universal gate sets for both classical and quantum computation, in which our definition of obfuscation can be met by polynomial-time algorithms. We also discuss some potential applications to testing quantum computers. We stress that the cryptographic security of these obfuscators, especially when composed with translation from other gate sets, remains an open question. 
\end{abstract}

\section{Introduction}

\subsection{Past work on circuit obfuscation}

Informally, an obfuscator is an algorithm that accepts a circuit as input, and outputs a hard-to-understand but functionally equivalent circuit. In this subsection, we briefly outline the state of current research in classical circuit obfuscation. To our knowledge, quantum circuit obfuscation has not been considered in any prior published work. 

Methods used for obfuscating logic circuits in practice have so far been essentially ad hoc~\cite{CT02, Simonaire}. Until recently, theoretical progress has primarily been in the form of no-go theorems for various strong notions of obfuscation~\cite{Barak, GR07}. The ability to efficiently obfuscate certain circuits would have important applications in cryptography. For instance, sufficiently strong obfuscation of circuits of the form ``encrypt with a hard-wired private key'' could turn a private-key encryption scheme into a public-key encryption scheme. As this example illustrates, one undesirable outcome is when the input circuit can be recovered completely from the obfuscated circuit. In this case, we say that the obfuscator \emph{completely failed} on that circuit~\cite{Barak}. Unfortunately, every obfuscator will completely fail on some circuits (e.g., learnable circuits.) On the other hand, there are trivial obfuscators which will erase at least some information from some circuits, e.g., by removing all instances of $X^{-1}X$ for some invertible gate $X$. 

In order to give a useful formal definition of obfuscation, one must decide on a reasonable definition of ``hard-to-understand.'' The most stringent definition in the literature demands \emph{black-box obfuscation}, i.e., that the output circuit is computationally no more useful than a black box that computes the same function. Barak et al.~\cite{BGIRSVY01} gave an explicit family of circuits that are not learnable and yet cannot be black-box obfuscated. They also showed that there exist (non-learnable) private-key encryption schemes that cannot be turned into a public-key cryptosystem by obfuscation. Their results do not preclude the possibility of black-box obfuscation for specific families of circuits, or of applying obfuscation to produce public-key systems from private ones in a non-generic fashion. It is an open problem whether quantum circuits can be black-box obfuscated.

A weaker but still quite natural notion is called \emph{best-possible obfuscation}; in this case, we ask that the obfuscated circuit reveals no more information than any other circuit that computes the same function. Goldwasser and Rothblum~\cite{GR07} showed that for efficient obfuscators, best-possible obfuscation is equivalent to \emph{indistinguishability obfuscation}, which is defined as follows. For any circuit $C$, let $|C|$ be the number of elementary gates, and let $f_C$ be the Boolean function that $C$ computes.

\begin{definition}\label{def:indistinguishability} A probabilistic
  algorithm $\mathcal O$ is an \emph{indistinguishability obfuscator}
  for the collection $\mathcal C$ of circuits if the following three
  conditions hold:
\begin{enumerate}
\item (functional equivalence) for every $C \in \mathcal C$, $f_{\mathcal O (C)} = f_C$;
\item (polynomial slowdown) there is a polynomial $p$ such that
  $|\mathcal O(C)| \leq p(|C|)$ for every $C \in \mathcal C$;
\item (indistinguishability obfuscation) For any $C_1, C_2 \in \mathcal C$ such that $f_{C_1} =
  f_{C_2}$ and $|C_1| = |C_2|$, the two distributions $\mathcal
  O(C_1)$ and $\mathcal O(C_2)$ are indistinguishable.
\end{enumerate}
\end{definition}

\noindent In the third part of the above definition, one must choose a notion of indistinguishability for probability distributions. Goldwasser and Rothblum~\cite{GR07} consider three such notions: perfect (exact equality), statistical (total variation distance bounded by a constant), and computational (no probabilistic polynomial-time Turing Machine can distinguish samples with better than negligible probability). They show that the existence of an efficient statistical indistinguishability obfuscator would result in a collapse of the polynomial hierarchy to the second level. This result also applies if the condition $|C_1| = |C_2|$ in property (3) of Definition~\ref{def:indistinguishability} is relaxed to $|C_1| = k|C_2|$ for any fixed constant $k$~\cite{GR07}.

A recent breakthrough has shown that computational indistinguishability may be achievable in polynomial time. Combining a new obfuscation scheme for NC1 circuits with fully homomorphic encryption, Sahai et al. gave an efficient obfuscator which achieves the computational indistinguishability condition under plausible hardness conjectures~\cite{GGHRSW13}. Subsequent work outlined a number of cryptographic applications of computational indistinguishability~\cite{SahaiW13}.

\subsection{Outline of present work}

\subsubsection{New notion of obfuscation}

An exact deterministic indistinguishability obfuscator would yield a solution to the circuit equivalence problem. For general Boolean circuits, this problem is co-NP hard. Therefore, exact deterministic indistinguishability obfuscation of general Boolean circuits cannot be achieved in polynomial time under the assumption $\mathrm{P} \neq \mathrm{NP}$. We propose an alternative route to weakening the exactness condition, by pursuing a notion of ``partial-indistinguishability''. In partial-indistinguishability obfuscation, we relax condition (3) so that it need only hold for $C_1$ and $C_2$ that are related by some fixed, finite set of relations on the underlying gate set.\footnote{Our construction for satisfying this definition uses \emph{reversible} gates. The definition of functional equivalence becomes more technical in that context, as discussed in Section \ref{sec:reversible}.}
 
\begin{definition}\label{def:partial-indistinguishability}
Let $G$ be a set of gates and $\Gamma$ a set of relations satisfied by
the elements of $G$. An algorithm $\mathcal O$ is a
\emph{$(G, \Gamma)$-indistinguishability obfuscator} for the
collection $\mathcal C$ of circuits over $G$ if the following three
conditions hold:
\begin{enumerate}
\item (functionality) for every $C \in \mathcal C$, $f_C = f_{\mathcal
    O(C)}$; 
\item (polynomial slowdown) there is a polynomial $p$ such that
  $|\mathcal O(C)| \leq p(|C|)$ for every $C \in \mathcal C$;
\item (($G, \Gamma$)-indistinguishability) for any $C_1,
  C_2 \in \mathcal C$ that differ by some sequence of applications
  of the relations in $\Gamma$, $\mathcal O(C_1) = \mathcal O(C_2)$.
\end{enumerate}
\end{definition}

The power of the obfuscation is now determined by the power of the relations $\Gamma$. If $\Gamma$ is a complete set of relations, generating all circuit equivalences over $G$, then a ($G, \Gamma$)-indistinguishability obfuscator is a perfect indistinguishability obfuscator according to Definition \ref{def:indistinguishability}. (Complete sets of relations for $\{\mathrm{Toffoli}\}$ and $\{\mathrm{AND},\mathrm{OR},\mathrm{NOT}\}$ are given in~\cite{Iwama, Huntington}.) If $\Gamma$ is the empty set then even the identity map fits the definition, and no obfuscation is taking place. With different sets of relations, one can interpolate between these extremes. The intermediate obfuscators form a partially ordered set, where a $(G, \Gamma')$-indistinguishability obfuscator is strictly stronger than a $(G, \Gamma)$-indistinguishability obfuscator if $\Gamma'$ is a strict superset of $\Gamma$. We remark that partial-indistinguishability is no stronger than perfect indistinguishability, and appears to be incomparable with statistical and computational indistinguishability. This is part of our motivation in considering this new definition. 

In the context of quantum computation, we make only a few minor changes to Definitions \ref{def:indistinguishability} and \ref{def:partial-indistinguishability}. First, the obfuscators will still be classical algorithms. On the other hand, the gates will be unitary and the circuits to be obfuscated will be unitary quantum circuits. Finally, the notion of functional equivalence now simply means that the operator-norm distance between the unitary implemented by $C$ and the unitary implemented by $\mathcal O(C)$ is bounded by a small constant $\epsilon > 0$.

\subsubsection{Group normal forms}

A finitely generated group can be specified by a presentation. This is a list of generators $\sigma_1,\ldots,\sigma_n$ and a list of relations obeyed by these generators. (A relation is simply an identity such as $\sigma_1 \sigma_3 = \sigma_3 \sigma_1$.) All group elements are obtained as products of the generators and their inverses. However, by applying the relations, we can get multiple words in the generators and their inverses that encode the same group element. A normal form specifies, for each group element, a unique decomposition as a product of generators and their inverses. For certain groups, including the braid groups, polynomial time algorithms are known which, given a product of generators and their inverses, can reduce it to a normal form. The word problem is, given two words in the alphabet $\{\sigma_1,\ldots,\sigma_n,\sigma_1^{-1},\ldots,\sigma_n\}$, to decide whether they specify the same group element. If a normal form can be computed, then this solves the word problem: just reduce both words to normal form and check whether the results are identical. However, an efficient solution for the word problem does not in general imply an efficiently computable normal form.

\subsubsection{Efficient constructions from group representations}

In this paper, we propose a general method of designing partial-indistinguishability obfuscators based on groups with efficiently computable normal forms. If a set of gates $G$ obeys the relations $\Gamma$ of the generators of a group with an efficiently computable normal form, then the reduction to normal form is an efficient $(G,\Gamma)$-indistinguishability obfuscator. The gates may obey additional relations beyond $\Gamma$, which is why the obfuscator does not solve the circuit-equivalence problem, which is believed to be intractable for both classical and quantum circuits. 

To demonstrate this method, we discuss an implementation using the braid groups $B_n$, for both classical reversible circuits and unitary quantum circuits. The number of strands $n$ in the braid group depends linearly on the number of dits or qudits on which the circuit acts. In Section \ref{sec:classical}, we describe a computationally universal reversible classical gate obeying the braid group relations, which was constructed in~\cite{Mochon, OP99, Kitaev03} from the quantum double of $A_5$. In Section \ref{fibonacci}, we describe a computationally universal quantum gate obeying the braid group relations, which was constructed in~\cite{FLW02} from the Fibonacci anyons. Our obfuscation scheme is similar in spirit to previously-proposed obfuscation schemes based on applying local circuit identities~\cite{Simonaire}, but the uniqueness of normal forms adds a qualitatively new feature. One consequence of this feature is that we can satisfy Definition \ref{def:partial-indistinguishability} and guarantee the partial-indistinguishability property against computationally unbounded adversaries. The running time of the obfuscator is the same as the running time of the the normal form algorithms, which take time $O(l^2 m \log m)$ for $m$-strand braids of length $l$~\cite{Dehornoy08}.

We remark that these gate sets that obey the braid group relations are not artificial constructions; in fact, they are the most natural choice in many contexts, some of which we list here. In the quantum case, these gates are native to certain proposed physical implementations of quantum computers~\cite{Kitaev03}, where the topological braiding property provides inherent fault-tolerance. The problem of approximating the Jones Polynomial invariant of links is complete for polynomial-time quantum computation~\cite{AA11}; an analogous fact is true for a restricted case of quantum computations motivated by NMR implementations~\cite{Shor_Jordan}. Both of these facts are naturally expressed in the gate set constructed from the Fibonacci representation. In the classical case, the gate set derived from quantum doubles of finite groups was recently used to show BPP-completeness for approximation of certain link invariants~\cite{Krovi_Russell}.

We remark that another potential group family for constructing partial-indistinguishability obfuscators are the mapping class groups MCG$(\Sigma_g)$ of unpunctured surfaces of genus $g$. These groups also have quantumly universal representations~\cite{Alagic} and an efficiently solvable word problem~\cite{Hamidi-Tehrani}. It is not known if there are also classically universal permutation representations, or if there are efficiently computable normal forms.

\subsubsection{Other gate sets}

In some applications the native gate set will be different than the
ones used in our construction. It is natural to ask if our obfuscators
can be used in these settings as well. By universality (quantum or
classical), one has an efficient algorithm $\braid$ which translates
circuits from the native gate set to the braiding gate set, as well as
an efficient algorithm $\circu$ for translation in the opposite
direction. We also let $\normf$ denote the
partial-indistinguishability obfuscator. One might then attempt to
obfuscate by applying the following:\\ \\

\begin{algorithm}\label{obfuscator}
\emph{\begin{enumerate}
\item input: a circuit $C$ on $n$ (qu)dits
\item output: The circuit $\circu( \normf( \braid(C)))$.
\end{enumerate}}
\end{algorithm}

We stress that, unlike the map $\normf$, the composed map $\normf
\circ \braid$ does not necessarily satisfy
Definition~\ref{def:partial-indistinguishability}. As we discuss in Section
\ref{compilation}, careless choice of the map $\braid$ can partially
or completely break the security of the obfuscator. Finding
translation algorithms securely composable with
partial-indistinguishability obfuscators is an area of current
investigation.  

\section{Relevant Properties of the Braid Group}\label{braid-groups}

\begin{figure}
\begin{center}
\includegraphics[width=0.5\textwidth]{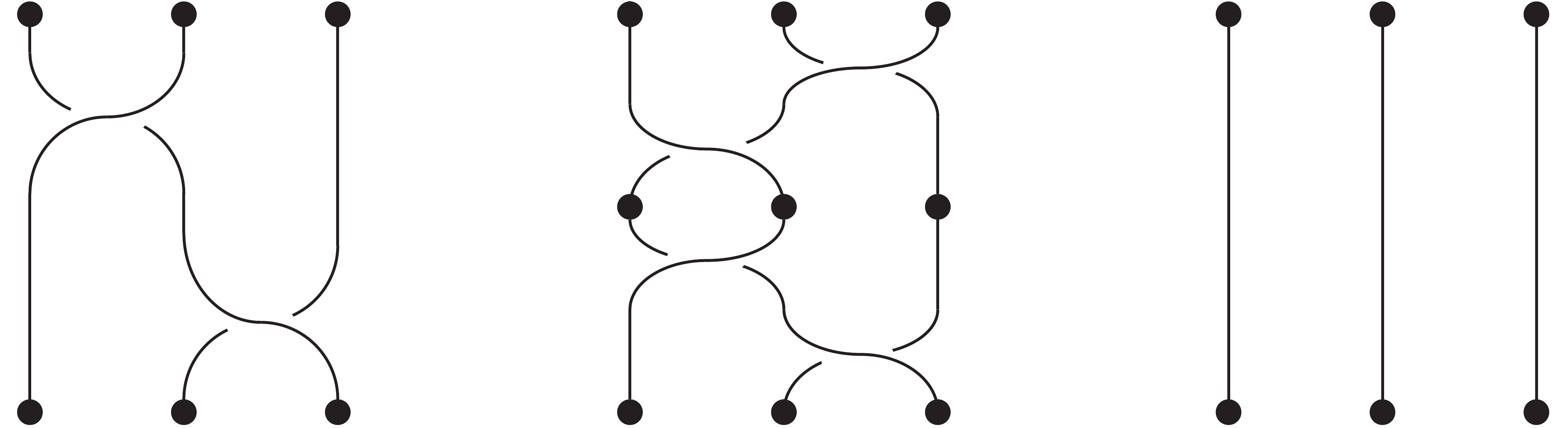}
\caption{\label{braids-example} The generator $\sigma_i$ represents
  the (clockwise) crossing of strands $i$ and $i+1$ connecting a
  bottom row of ``pegs'' to a top row. Multiplication of
  group elements corresponds to composition of braids. As an example,
  we show the 3-strand braid  $\sigma_1^{-1}\sigma_2$ (left), and the
  same braid composed with its inverse $\sigma_2^{-1}\sigma_1$
  (middle), which is equivalent to the identity element of $B_3$
  (right).}
\end{center}
\end{figure}

The braid group $B_n$ is the infinite discrete group with generators
$\sigma_1, \ldots, \sigma_{n-1}$ and relations
\begin{equation}\label{abstract-braids}
\begin{array}{rcll}
\sigma_i \sigma_j & = & \sigma_j \sigma_i & \forall \ |i-j| \geq 2\\
\sigma_i \sigma_{i+1} \sigma_i & = & \sigma_{i+1} \sigma_i \sigma_{i+1}
& \forall \ i.
\end{array}
\end{equation}
The group $B_n$ is thus the set of all words in the alphabet
$\{\sigma_1,\ldots,\sigma_{n-1},\sigma_1^{-1},\ldots,\sigma_{n-1}^{-1}\}$,
up to equivalence determined by the above relations. In 1925 Artin
proved that the abstract group defined above precisely captures the
topological equivalence of braided strings~\cite{Artin25}, as
illustrated in Fig. \ref{braids-example}. A charming exposition of
this subject can be found in~\cite{Kauffman91}.

In the word problem on $B_n$, we are given words $w$ and $z$, and our
goal is to determine if they are equal as elements of $B_n$. One
solution is to put both $w$ and $z$ into a \emph{normal form},
and then check if they are equal as words. For our purposes, it is
enough to describe the normal form and specify the complexity of
the algorithm for computing it. The details of the algorithm, along
with a thorough and accessible presentation of the relevant facts
about braids, can be found in~\cite{Dehornoy08}.

We first observe that the word problem is easily shown to be decidable if we
restrict our attention to an important subset of $B_n$. Note that the
presentation \eqref{abstract-braids} can also be viewed as a
presentation of a monoid, which we denote by $B_n^+$. The
elements of $B_n^+$ are called \emph{positive braids}, and are words
in the generators $\sigma_i$ only (no inverses), up to equivalence
determined by the relations in \eqref{abstract-braids}. Since all the
relations of $B_n$ preserve word length, and there are only finitely
many words of any given length, we can decide the word problem (albeit
very inefficiently) simply by trying all possible combinations of the
relations.

Building upon this, one can give an (inefficient) algorithm for the
word problem on $B_n$ itself~\cite{Gonzalez10}. First, given two
elements $a, b$ of $B_n^+$, we write $a \preccurlyeq b$ if there
exists $z \in B_n^+$ such that $b = az$; in this case we say
that $a$ is a \emph{left divisor} of $b$. Similarly, we write $a
\succcurlyeq b$ if there exists $y \in B_n^+$ such that $b = ya$; in
this case we say that $a$ is a \emph{right divisor}\footnote{The
  terminology is not accidental; it turns out that we can also define
  l.c.m.s and g.c.d.s in $B_n^+$, and that $B_n$ is the group of
  fractions of $B_n^+$. These facts are some of the achievements of
  Garside theory~\cite{Garside69}.} of $b$. The center of $B_n$ is the
cyclic group generated by $\Delta_n^2$, where
$$
\Delta_n := \Delta_{n-1}\sigma_{n-1} \sigma_{n-2} \cdots \sigma_1 \in B_n^+
$$
(see p.30 of~\cite{Gonzalez10} for a simple proof). Geometrically,
$\Delta_n$ implements a twist by $\pi$ in the $z$-plane as the strands
move from $z=0$ to $z=1$. One can show that $\sigma_i \preccurlyeq
\Delta_n$ for all $i$, i.e. there exists $x_i \in B_n^+$ such that
$\sigma_i^{-1} = x_i\Delta_n^{-1}$. Given a word $w$ in the $\sigma_i$
and their inverses, we first replace the leftmost instance of an inverse
generator (say it is $\sigma_i^{-1}$) with $x_i \Delta_n^{-1}$. We
then insert $\Delta_n^{-1} \Delta_n$ in front of $x_i$, and observe
that conjugating a positive braid $x$ by $\Delta_n$ results in another
positive braid (specifically, the rotation of $x$ by $\pi$ in the
$z$-plane). In this way, we can push $\Delta_n^{-1}$ all the way to
the left. We repeat this process for each inverse generator appearing
in the word, resulting in a word of the form $\Delta_n^p b$ where $p
\in \mathbb{Z}$ and $b \in B_n^+$. Since we can solve the word problem in
$B_n^+$, we can factor out the maximal power of $\Delta_n$ appearing
as a left divisor of $b$. We thus have that, as elements of the braid
group, $w = \Delta_n^{p'}b'$ with $\Delta_n$ not a left divisor of
$b'$ and $p'$ unique. This solves the word problem in $B_n$.

We can make the above algorithm efficient by finding an efficiently
computable normal form for a positive braid word $b$ that does not
have $\Delta_n$ as a left divisor. Recall that the symmetric group
$S_n$ has a remarkably similar presentation to $B_n$. Indeed, starting
with \eqref{abstract-braids}, letting $\sigma_i = (i~i+1)$ and adding
the relations $\sigma_i^2 = 1$ for all $i$ results in the standard
presentation of $S_n$. In other words, there is a surjective
homomorphism $\pi: B_n \rightarrow S_n$ with $\sigma_i \mapsto
(i~i+1)$. In terms of the geometric interpretation, a braid is mapped
to the permutation on $[n]$ defined by the connections between the
top and bottom ``pegs,'' as in Figure \ref{braids-example}. For each $\sigma\in S_n$, there is a unique preimage of $\sigma$ that can be drawn so that any given pair of strands cross only in the positive direction, and at most once. We call such braids \emph{simple braids}, and they form a subset of $B_n^+$ of size $n!$. 

\begin{definition}\emph{p.4 of \cite{Dehornoy08}.}
\begin{enumerate}
\item A sequence of simple braids $(s_1, \dots, s_p)$ is said to be
\emph{normal} if, for each $j$, every $\sigma_i$ that is a left
divisor of $s_{j+1}$ is a right divisor of $s_j$.
\item A sequence of permutations $(f_1, \dots, f_p)$ is said to be
\emph{normal} if, for each $j$, $f_{j+1}^{-1}(i) > f_{j+1}^{-1}(i+1)$
implies $f_j(i) > f_j(i+1)$.
\end{enumerate}
\end{definition}

\noindent A sequence of simple braids $(s_1, \dots, s_p)$ is normal if
and only if the sequence of permutations $(\pi(s_1), \dots, \pi(s_p))$
is normal. Given a permutation $f \in S_n$, let $\hat f$ denote the
simple braid of $B_n$ satisfying $\pi (\hat f) = f$.

\begin{theorem}\emph{p.4 of \cite{Dehornoy08} and Ch.9 of
    \cite{Epstein92}.}\label{thm:normal-form}
\begin{enumerate}
\item Every braid $z$ in $B_n$ admits a unique decomposition of the
form $\Delta_n^m s_1 \dots s_p$ with $m \in \mathbb{Z}$ and $(s_1,
\dots, s_p)$ a normal sequence of simple braids satisfying $s_1 \neq
\Delta_n$ and $s_p \neq 1$.
\item Every braid $z$ in $B_n$ admits a unique decomposition of the
form $\Delta_n^m \hat f_1 \dots \hat f_p$ with $m \in \mathbb{Z}$ and
$(f_1, \dots, f_p)$ a normal sequence of permutations satisfying $f_1
\neq \pi(\Delta_n)$ and $f_p \neq 1$.
\end{enumerate}
\end{theorem}

\noindent The most efficient algorithms for computing the normal form
of a word $w$ in the generators of $B_n$ have complexity $O(|w|^2 n
\log n)$~\cite{Dehornoy08}.

\section{Obfuscation of Classical Reversible Circuits}
\label{sec:classical}

\subsection{Reversible Circuits}
\label{sec:reversible}

In the next section, we will describe a gate $R$ which is universal for classical computation and satisfies Definition \ref{def:partial-indistinguishability} when $\Gamma$ is the set of relations of the braid group. Because group elements are invertible, $R$ must be a reversible gate, that is, it must bijectively map its possible inputs to its possible outputs. We will thus work in the setting of \emph{reversible classical circuits}. These circuits are composed entirely of reversible gates. For more background on reversible computation see~\cite{Bennett, Fredkin_Toffoli, Nielsen_Chuang}.

Because reversible circuits cannot erase any information, they operate using ancillary dits (``ancillas'') to store unerasable data left over from intermediate steps in the computation. A reversible circuit evaluating a function $f:\{0,\ldots,d-1\}^n \to \{0,\ldots,d-1\}^m$ thus operates on $r \geq \max(n,m)$ dits, where $r-n$ of the input dits are work dits to be initialized to some fixed value independent of the problem instance, and $r-m$ of the output dits contain unerasable leftover data, to be ignored. Efficient procedures are known for compiling arbitrary logic circuits into reversible form, e.g., by using the Toffoli (or CCNOT) gate~\cite{Bennett, Fredkin_Toffoli}.

In adapting Definitions \ref{def:indistinguishability} and \ref{def:partial-indistinguishability} to reversible circuits, one is faced with two natural choices for the notion of functional equivalence. One may either demand that the original and obfuscated circuits implement the same function $f:\{0,1\}^n \to \{0,1\}^m$, ignoring the ancilla dits (\emph{weak equivalence}), or demand that they implement the same transformation on the entire set of $r$ dits,
including the ancillas (\emph{strong equivalence}). Our constructions will satisfy the latter. Strong equivalence implies weak equivalence, so our construction proves that both possible definitions of partial-indistinguishability are polynomial-time achievable when $\Gamma$ is the set of relations of the braid group. We remark that, as with ordinary irreversible circuits, determining if two arbitrary reversible circuits are equivalent (weakly or strongly) is coNP-complete~\cite{Jordan13}.

\subsection{Classical computation with braids}\label{sec:universal}

We now briefly describe a classical reversible gate $R$ which satisfies the braid relations. The complete details of the construction and the proof of universality of $R$ are given in Appendix \ref{universal}. Taken together with Theorem \ref{thm:normal-form}, this yields an obfuscator satisfying Definition \ref{def:partial-indistinguishability}. 

Let $G$ be a finite group and set $d = |G|$. Consider the reversible gate $R$ that acts on pairs of dits encoding group elements by
\begin{equation}\label{qdouble}
R(a,b) = (b, b^{-1}ab).
\end{equation}
Let $R_i$ denote $R$ acting on the $i$ and $(i+1)^{\mathrm{th}}$ wires of a circuit. By direct calculation, one can check that the set $\{R_1,\dots,R_{n-1}\}$ satisfies the braid relations, that is,
\begin{equation}
\begin{array}{rcll}
R_i R_j & = & R_j R_i & \forall \ |i-j| \geq 2\\
R_i R_{i+1} R_i & = & R_{i+1} R_i R_{i+1}
& \forall \ i.
\end{array}
\end{equation}
In 1997, Kitaev discovered that the gate set $\{R, R^{-1}\}$ is universal for classical reversible computation when $G$ is the symmetric group $S_5$~\cite{Kitaev03}. Ogburn and Preskill subsequently showed that the alternating group $A_5$, which is half as large as $S_5$, is already sufficient~\cite{OP99}. The universality construction for $A_5$ was subsequently presented in greater detail and generalized to all non-solvable groups by Mochon~\cite{Mochon}. To make our presentation more accessible and self-contained, we give in Appendix \ref{universal} an explicit description of Mochon's universality construction in the the case $G = A_5$. The construction proves computational universality by showing how to efficiently compile Toffoli circuits into $R$-circuits. 

Given any $R$-circuit, we can apply the algorithm of Theorem \ref{thm:normal-form} by interpreting each $R_i$ as $\sigma_i$ and each $R_i^{-1}$ as $\sigma_i^{-1}$. This leads to partial-indistinguishability obfuscation of $R$-circuits. A discussion of whether this can also yield meaningful obfuscation for classical circuits constructed from other gate sets is given in Section \ref{attacks}.

\section{Quantum Circuits}\label{sec:quantum}

\subsection{Quantum computation with braids}
\label{fibonacci}

In Section \ref{sec:universal} and Appendix \ref{universal}, we discuss classical universality of circuits encoded as braids. It turns out that an analogous theory can be developed for quantum circuits, and is well-understood. The family of so-called Fibonacci representations of the braid groups have dense image in the unitary group, and there are efficient classical algorithms for translating any quantum circuit into a braid (and vice-versa) in a way that preserves unitary
functionality~\cite{FLW02}. A brief synopsis of these facts is given below. We remark that there are in fact many unitary representations of the braid groups that satisfy these properties, and which are physically motivated by the so-called fractional quantum Hall effect. In this setting, the image of these representations consists of unitary operators which describe the braiding of excitations in a 2-dimensional medium~\cite{Kitaev03}.

Approachable descriptions of the Fibonacci
representation are given in~\cite{Shor_Jordan, Trebst}. In
\cite{Shor_Jordan}, what we call the ``Fibonacci representation''
here, is called the ``$\star \star$'' irreducible
sub-representation. This is a family of representations
$\rho^{(n)}_{\mathrm{Fib}}:B_n \rightarrow U(F_{n-4})$, where $F_k$ is the
$k$-th Fibonacci number. For our application, the essential properties
of the Fibonacci representation are \emph{locality} and \emph{local
  density}. These two properties mean that, under a certain qubit
encoding, braid generators correspond to local unitaries, and local
unitaries correspond to short braid words. Standard arguments from
quantum computation tell us that we can achieve the latter to
precision $\epsilon$ with $O(\log^{2.71}(1/\epsilon))$ braid generators by
means of the Solovay-Kitaev algorithm~\cite{Dawson_Nielsen}. 

A natural basis for the space of $\rho^{(n)}_{\mathrm{Fib}}$ can be identified
with strings of length $n$ from the alphabet $\{\star, p\}$, which
begin with $\star$, end with $p$, and do not contain ``$\star \star$''
as a substring\footnote{In \cite{Shor_Jordan} the $\star \star$
  subrepresentation of $B_n$ acts on strings of length $n+1$ that
  begin and end with $\star$. One can leave the initial and/or final
  $\star$ implicit as these are left unchanged by all braiding operations. We
  omit the final $\star$ leaving us strings of length $n$ that begin
  with $\star$ and end with $p$.}. Following
\cite{AA11}\footnote{Reference \cite{AA11} describes
  the basis vectors in terms of ``paths''. The correspondence between
  the path notation and the $p \star$ notation is given in appendix C
  of~\cite{Shor_Jordan}.}, for $n$ a multiple of four, we identify a
particular subspace $V_n$ of $\rho^{(n)}_{\mathrm{Fib}}$ by discarding some
basis elements, as follows. Partition a string $s$ into substrings of
length four. If each of these substrings is equal to either $\star p
\star p$ (this will encode a $0$) or $\star p p p$ (this will encode a
$1$), then the basis element corresponding to $s$ is in $V_n$;
otherwise, it is not. Note that $\dim V_n = 2^{n/4}.$ The following
theorem follows from \cite{AA11, Dawson_Nielsen}.

\begin{theorem}
\label{unifib}
There is a classical algorithm which, given an $(n/4)$-qubit quantum
circuit $C$ and $\epsilon>0$, outputs a braid $b \in B_n$ of length
$O(|C|\log^{2.71}(1/\epsilon))$ satisfying
$$
\left\|C - \left.\rho^{(n)}_{\mathrm{Fib}}(b)\right|_{V_n}\right\| \leq \epsilon~;
$$
this algorithm has complexity $O(|b|)$.
\end{theorem}

\noindent For the opposite direction, we can identify a subspace $W_n
\subset (\mathbb C_2)^{\otimes n}$ by discarding all bitstrings except
those that start with $0$, end with $1$ and do not have ``$00$'' as a
substring. Then $\dim W_n = \dim \rho^{(n)}_{\mathrm{Fib}}$ and we have the
following.

\begin{theorem}
There is a classical algorithm which, given $b \in B_n$ and
$\epsilon>0$, outputs a quantum circuit $C$ on $n$ qubits of length
$O(|b|\log^{2.71}(1/\epsilon))$ such that
$$
\left\|\left.C\right|_{W_n} - \rho^{(n)}_{\mathrm{Fib}}(b)\right\| \leq \epsilon~;
$$
this algorithm has complexity $O(|C|)$.
\end{theorem}
\noindent The two algorithms in the above theorems are described
explicitly in~\cite{AA11}. 

\subsection{Obfuscating quantum computations}

While the state of knowledge about classical obfuscation is limited, essentially nothing is known about the quantum case. Here we discuss how to use the facts from the previous section to construct a partial-indistinguishability obfuscator for quantum circuits. 

In light of Theorem \ref{unifib}, $\{\rho_{\mathrm{Fib}}(\sigma_1),\ldots,\rho_{\mathrm{Fib}}(\sigma_{n-1})\}$ may be regarded as a universal set of elementary quantum gates. By the homomorphism property of $\rho_{\mathrm{Fib}}$, this set satisfies the braid relations. These gates differ from conventional quantum gates in that they do not possess locality defined in terms of a strict tensor product structure. Nevertheless, as shown above, the power of unitary circuits composed from these gates is equivalent to standard quantum computation. By interpreting each $\rho_{\mathrm{Fib}}(\sigma_j)$ as a braid-group generator $\sigma_j$, we can apply the algorithm from Theorem \ref{thm:normal-form} directly to circuits from this gate set, resulting in a partial-indistinguishability obfuscator satisfying Definition \ref{def:partial-indistinguishability}.

With the algorithms from the previous section in hand, we could also attempt to apply the obfuscation algorithm, Algorithm \ref{obfuscator}, directly to quantum circuits. For an input circuit $C$ on $n$ qubits, the running times of both of this algorithm is $O(|C|^2 n \cdot \text{polylog}(n, 1/\epsilon))$. The length of the output cannot be longer than the running time. We are not aware of a better upper bound for the length of the output. The security of this algorithms is questionable, and some attacks and possible countermeasures are discussed in Section \ref{attacks}.

Note that reduction of arbitrary quantum circuits to a normal form using a \emph{complete} set of gate relations should not be possible in polynomial time; this would yield a polynomial-time algorithm for deciding whether a quantum circuit is equivalent to the identity, which is a coQMA-complete problem~\cite{JWB03}.

\subsection{Testing claimed quantum computers with a quantum obfuscator}

It is natural to consider quantum analogues of the applications of
obfuscation from classical computer science. We now consider a potential
application of quantum circuit obfuscation that does not fit this
mold: testing claimed quantum computers. A similar proposal using a
restricted class of quantum circuits has been previously made in
\cite{SB09}. 

Suppose Bob claims to have 
access to a universal quantum computer with some fixed finite number
of qubits. Alice has access to a classical computer only, as well as a
classical communication channel with Bob. Can Alice determine if Bob
is telling the truth? Barring tremendous advances in complexity
theory, a provably correct test is unlikely;\footnote{Notice that even
  a proof that BQP $\neq$ BPP would be insufficient; one would have to
  find specific problems and instance sizes where some quantum
  strategy provably beats every classical one. We are thus left with a
  situation analogous to the practical security guarantees of modern
  cryptographic systems, which tell us how many bit operations it
  would take to crack a given instance using the fastest known
  algorithms.} can we still design a test in which we have a high
degree of confidence? Given the extensive work on classical algorithms
for factoring, a reasonable idea is to simply ask Bob to factor a
sufficiently large RSA number. However, Shor's algorithm only begins
to outperform the best classical algorithms when thousands of logical
qubits can be employed. A much smaller universal quantum computer
(e.g., a few dozen qubits) is likely to be a far simpler engineering
challenge and could still be quite useful, e.g., for simulating
certain quantum systems. A test that works in this case would thus be
very valuable. We now outline a new proposal for such a test. Simply
put, we propose asking questions that are classically easy to answer,
but posing them in an obfuscated manner. In this test,  Alice would
repeatedly generate quantum circuits and ask Bob to run them. At least
some of the circuits would in fact be quantumly-obfuscated classical
reversible circuits, allowing Alice to easily check the
answers. Previous work has yielded tests of quantum computers in the
case that the verifier can perform some limited quantum
operations~\cite{BFK08, ABE08}.

We have considerable freedom when designing an obfuscation-based test
of quantum computers. How to choose these parameters in a way that
makes the test difficult to fool with a classical computer is an open
question. For purposes of illustration, we give one example. Let
$\mathcal O$ be the obfuscation algorithm for quantum circuits
described above.

\begin{algorithm}
\emph{\begin{enumerate}
\item Select a random bitstring $s$ of length $k$.
\item Let $C$ be the $(k+1)$-bit circuit that, on all-zero input, initializes wires $2$ through $k+1$ to $s$ and then computes the parity of $s$ into the first wire.
\item Compute $\mathcal O(C)$, and let $n$ be the number of qubits needed to run $\mathcal O(C)$.
\item Ask Bob to run $D$ on the all zeros string and return the first
  bit of output.
\end{enumerate}}
\end{algorithm}

\noindent Clearly, $k$ must be chosen so that $n$ is smaller than the
number of logical qubits Bob claims to control. To fool Alice, a
purely classical Bob must determine the parity of $s$. The dictionary
attack (\emph{i.e.} Bob repeatedly guesses at $k$, obfuscates the corresponding
circuit, and compares the result to the circuit given by Alice) is of no use
provided $k$ is reasonably large, e.g., 80 bits, which can be encoded
using a braid of 115 strands using the Zeckendorf encoding described
in~\cite{Shor_Jordan}.

We now show that there can be no efficient general-purpose algorithm
for breaking our test by detecting whether a given quantum circuit is
in fact (almost) classical, and if so, simulating it.

\begin{definition}
\label{def:class}
Let $c$ be a bit string specifying a quantum circuit via a standard
universal set $Q$ of quantum gates, and let $U_c$ be the corresponding
unitary operator. Fix some constants $r,d,a \in \mathbb{N}$, and fix a
set $R$ of reversible gates. The problem
$\mathrm{CLASS}(r,d,a,Q,R)$ is to find a reversible circuit of
at most $r |c|^d$ gates from $R$ such that the corresponding
permutation matrix $P$ satisfies $\|U_c - P \| \leq 2^{-a |c|}$.
\end{definition}

\noindent Note that $\mathrm{CLASS}(r,d,a,Q,R)$ is not a decision
problem. Thus, to formulate the question of whether this problem can
be efficiently solved, we must ask not whether
$\mathrm{CLASS}(r,d,a,Q,R)$ is contained in P but whether it is
contained in FP. We now provide some formal evidence that this is not
the case. Note that the following theorems continue to hold if we
change the classicality condition in Definition~\ref{def:class} to $\|U_c - P \|
\leq |c|^{-a}$.

\begin{theorem}\label{th:testing}
For any fixed $r,d,a \in \mathbb{N}$, any universal reversible
gate set $R$, and any universal quantum gate set $Q$, if 
$\mathrm{CLASS}(r,d,a,Q,R) \in \mathrm{FP}$
then $\mathrm{QCMA} \subseteq \mathrm{P}^{\mathrm{NP}}$.
\end{theorem}

\noindent Note that, $\mathrm{QCMA} \subseteq \mathrm{P}^{\mathrm{NP}}$ would be
very surprising because, among other things, it would imply
$\mathrm{BQP} \subseteq \mathrm{PH}$, and there is evidence that this
is false \cite{Aaronson, Fefferman}.

\begin{proof}
The standard QCMA-complete language $\mathcal{L}$ is as follows. Let
$\mathcal C$ be the set of all quantum circuits (expressed as a
concatenation of bitstrings that index elements of the gate set $Q$). $\mathcal C$
decomposes as the disjoint union of $\mathcal{L}$ and
$\bar{\mathcal{L}}$ where $\mathcal{L}$ consists of the quantum
circuits that accept at least one classical (\emph{i.e.} computational
basis state) input, and $\bar{\mathcal{L}}$ consists of the circuits
that reject all inputs. Given a quantum circuit $V_1 \in \mathcal C$,
(the ``verifier'') we can amplify it using standard
techniques~\cite{Marriott_Watrous, Nagaj_Wocjan_Zhang} to
accept YES instances with probability at least $1-O(2^{-n})$ and
accept NO instances with probability at most $O(2^{-n})$. Let $V_2$ be
such an amplified verifier. Further, let
\[
\begin{array}{lcr}
V_3 & = &
\begin{array}{c}
\Qcircuit @C=1em @R=.5em {
& \qw                & \targ     & \qw               & \qw \\
& \multigate{1}{V_2} & \ctrl{-1} & \multigate{1}{V_2^{-1}} & \qw \\
& \ghost{V_2}        & \qw       & \ghost{V_2^{-1}}        & \qw  
}
\end{array}
\end{array}
\]
where the second-to-top qubit is the acceptance qubit of $V_2$. If
$V_i \in \bar{\mathcal{L}}$ then $\| V_3 - \id \| = O(2^{-n})$. By
assumption, there exists a polynomial time classical algorithm for
solving $\mathrm{CLASS}(r,d,a,Q,R)$. When presented with
$V_3$, this algorithm will produce a polynomial-size reversible
circuit $V_4$ strongly equivalent to the identity. By querying an
oracle for the problem of strong equivalence of reversible circuits,
one can decide whether $V_4$ is equivalent to the circuit of no gates,
and hence to the identity operation. If $V_1 \in \bar{\mathcal{L}}$,
this oracle will accept. If $V_1 \in \mathcal{L}$ then the
algorithm for problem 1 will answer NO or produce a circuit that this
oracle rejects. As shown in~\cite{Jordan13}, the problem of
deciding strong equivalence of reversible circuits is contained in
coNP. Thus, we can decide QCMA in $\mathrm{P}^{\mathrm{coNP}}$, which
is equal to the more familiar complexity class $\mathrm{P}^{\mathrm{NP}}$.
\end{proof}

\section{Some Attacks}
\label{attacks}

\subsection{Compiler attacks}
\label{compilation}

The security or insecurity of braid-based partial-indistinguishability
obfuscation remains an area of current investigation. From a purely
information-theoretic point of view, the power of this obfuscation comes from the
many-to-one nature of the map $\normf$ that takes arbitrary braid words to
their normal form. If the initial braid words are highly structured
because they are obtained by compilation from a different gate set,
then this can undermine or destroy the many-to-one feature of
$\normf$.

In Section \ref{sec:universal}, %Appendix \ref{universal}, 
we describe a reversible gate $R$ on
pairs of 60-state dits, corresponding to elements of $A_5$, that obeys the relations of the braid group and
can perform universal classical computation. The gate itself and the
proof that it is universal come from the quantum computation
literature~\cite{Kitaev03, OP99, Mochon}. Appendix \ref{universal}
recounts the universality proof of \cite{Mochon}, which can be viewed
as a compiler $\braid_{R}$ that maps circuits constructed from the
well-known universal reversible Toffoli gate into circuits constructed
from the $R$ gate. As a cautionary example, we now show that naively
obfuscating Toffoli circuits using the composed map $\normf \circ
\braid_{R}$ is completely insecure.

The construction in Appendix \ref{universal} gives a general mapping from a Toffoli gate to a corresponding braid. We will refer to braids obtained in this way as \emph{Toffoli braids}. Recall that the normal form of a braid in $B_n$ has the form $\Delta_n^m s_1\dots s_p$ for a normal sequence of simple braids $(s_1,\dots,s_p)$. A Toffoli braid obtained from a Toffoli with controls $c_1$ and $c_2$ and target $t$ has normal form 
\begin{equation}\label{eq:tof-braid}
\Delta_n^0 s_1(c_1,c_2,t)s_2s_3s_4s_5s_6s_7s_8s_9(c_1,c_2,t)s_{10}s_{11}s_{12}s_{13}(c_1,c_2,t)s_{14}(t).
\end{equation}
The factors $s_2,\dots,s_8,s_{10},s_{11}$ and $s_{12}$ only depend on $n$, and not on the wires $c_1$, $c_2$ or $t$. Note that this is a positive braid --- consisting only of $\sigma_1,\dots,\sigma_{n-1}$ and none of their inverses. Any product of such braids will thus also be a positive braid, so attempting to obfuscate a circuit in Toffoli gates using this construction will yield only positive braids. 

Because Toffoli is a 3-bit gate, there are only $\binom{n}{3}$ ways to
apply a Toffoli to $n$ bits. Thus, one may, in polynomial time,
test each of these $\binom{n}{3}$ possibilities as a guess for the
last gate of the obfuscated circuit. One performs the test by
compiling the guessed Toffoli gate into a braid, appending the inverse
of this braid to the normal form braid produced as the output the
obfuscator, and then reducing the resulting braid to normal form. If
the guess is correct, then the resulting braid is still a braid corresponding to a circuit --- the original obfuscated circuit with its last Toffoli gate removed --- and thus this will result in a
positive braid. If the guess is incorrect, then appending the inverse
of a positive braid, which consists entirely of
$\sigma_1^{-1},\dots,\sigma_{n-1}^{-1}$, might result in a braid that
is no longer positive --- that is, has a negative power of $\Delta_n$,
and this seems to be the case with \emph{any} wrong guess, based on
some limited tests. Furthermore, the presence of a negative power of
$\Delta_n$ is efficiently recognizable, so it is immediately clear
whether or not the guess was correct.

This attack is related to so-called length-based
attacks. These have been introduced in the cryptanalysis of braid
based key-exchanged protocols~\cite{HS03}. In the present context, the
natural length-based attack is to guess the final gate, append the
inverse of the corresponding braid to the normal-form braid produced
by the obfuscator, and the reduce the product braid to normal form. If
the result is a shorter word in the braid-group generators
than the original normal form, then this can be taken as heuristic
evidence that the guess was correct. Intuitively, one expects that the
longer the braid words are that implement individual gates from the
original gate set, then the better such attacks should work.

One can easily propose modifications to the naive obfuscator 
$\normf \circ \braid_{R}$ that thwart guessing-based attacks such as
the two attacks described above. In particular, one finds that the
gate $R$ described in Appendix \ref{universal} has order $60$. Hence,
one can start with the %124-generator 
positive Toffoli braid in equation \eqref{eq:tof-braid} and
then each generator $\sigma_i$ can independently, with probability
$\frac{1}{2}$, be replaced with $\sigma_i^{-59}$, without altering the
functionality of the circuit. The number of generators in a Toffoli braid depends on $n$, and which wires the Toffoli acts on, but there are always at least 124. Thus, each gate will be compiled into
one of $2^{124}$ braid-words uniformly at random. Thus, guessing-based
attacks on the composition of this compiler with $\normf$ may become
impractical. Whether such a scheme is vulnerable to other attacks
remains an open question for future research.

\subsection{Dictionary attacks}
\label{dictionary}

The partial-indistinguishability obfuscator described in the preceding
sections deterministically maps input circuits to obfuscated
circuits. This creates a potential weakness in the
obfuscation. Suppose Alice wishes to run a computation $C$ on Bob's
server but does not wish Bob to know what computation she is
running. Thus, she sends the obfuscated circuit $\mathcal{O}(C)$ to
Bob, who executes it, and returns the result. To improve security,
Alice may instead use a circuit $C'$ in which her desired input is
hard-coded, and which applies a one-time pad at the end of the
computation. If the obfuscation is secure, then Bob is unlikely to
learn anything about $C$, the input, or the output. However, if Bob
knows that the circuits Alice is likely to want to execute are drawn
from some small set $S$, then Bob can simply compute
$\{\mathcal{O}(s)|s \in S\}$ and identify Alice's computation by
finding it in this list. Such attacks are sometimes called
``dictionary'' attacks after the practice of recovering passwords by
feeding all words from a dictionary into the hash function and
comparing against the hashed password.

Dictionary attacks may or may not be a serious threat to our
obfuscation scheme, depending on the the size of the set of likely
circuits to be obfuscated. In cryptographic applications where
dictionary attacks are a concern, the standard way to protect against
them is to append  random bits prior to encryption. (In the context of
hashing passwords, this practice is called ``salting''.) Such a
strategy can be applied to our obfuscator, but some care must be taken
in doing so. The most obvious strategy is to append a random circuit
on the output ancillas prior to obfuscation. However, attackers can
defeat this countermeasure by using the polynomial-time algorithms for
computing left-greatest-common-divisors in the braid group
\cite{Epstein92}. However, prior to obfuscation, one may introduce
extra dits, and apply random circuits before, after, and simultaneously
with the computation, in a way so as not to disrupt it. The problem of
optimizing the details of this procedure so as to maximize security
and efficiency is left to future work.

\section{Future Work}

\subsection{Classical and quantum universality}

It is of interest to consider other computationally universal
representations of the braid group, which might provide more efficient
translations from circuits to braids. One avenue for obtaining such
representations is by finding other solutions to the Yang-Baxter
equation, besides the operator $R$ from Appendix \ref{universal}. Our
investigations so far prove that no permutation matrix solution of
dimension up to $16 \times 16$ is a universal gate and suggest that no
permutation matrix solution of dimension $25 \times 25$ is a universal
gate. In the quantum case, it has been shown that no $4 \times 4$
unitary solution is universal~\cite{ABJ12}.

More generally, one may look for other finitely-generated groups with
computationally universal representations and efficiently computable
normal forms. One potential candidate family are the mapping class
groups MCG$(\Sigma_g)$ of unpunctured surfaces of genus $g$. These
groups also have quantumly universal representations~\cite{Alagic} and
an efficiently solvable word problem~\cite{Hamidi-Tehrani}. It is not
known if there are also classically universal permutation
representations, or if there are efficiently computable normal forms.

\subsection{Expanding the set of indistinguishability relations}

By \cite{Jordan13}, achieving efficient indistinguishability
obfuscation for the complete set of relations of a universal gate set
is unlikely. However, it is possible that partial-indistinguishability
obfuscation on $R$ gates could be achieved with a larger set of
relations than the braid relations. For example, the universal
reversible gate described in Appendix \ref{universal} has order 60. If
we add the relations $\sigma_i^{60} = \id$ for $i=1,2,\ldots,n-1$ to
$B_n$, we obtain a ``truncated'' (but still infinite for large
$n$~\cite{Coxeter}) factor of the braid group. If a normal form can
still be computed in polynomial time for this group then one could
construct an efficient obfuscator using the relations of this
truncated group, which would be strictly stronger than our braid group
obfuscator. This approach also provides motivation for finding a
complete set of relations for the gate $R$. 

\section*{Acknowledgements}

We thank Anne Broadbent, Rainer Steinwandt, Scott Aaronson, Bill
Fefferman, Leonard Schulman, Robert K\"onig, and Yi-Kai Liu for helpful
discussions. We also thank Mariano Su\'arez-Alvarez and Gjergji Zaimi
for leading us to reference \cite{Coxeter} via \texttt{math.stackexchange} and
\texttt{mathoverflow}. Portions of this paper are a contribution of
NIST, an agency of the US government, and are not subject to US copyright.

\appendix

\section{Classical Computation with Braids}
\label{universal}

In this section, we present a reversible gate $R$ on pairs of 60-state
dits that can perform universal computation and obeys the relations of
the braid group. The universality construction for this gate comes
from the quantum computation literature~\cite{Kitaev03, OP99, Mochon},
but we present it here in purely classical language to make it
accessible to a broader audience.

\begin{figure}[h]
\centerline{
\Qcircuit @C=1em @R=.5em {
& \multigate{1}{R_1} & \qw                & \qw                & \qw\\
& \ghost{R_1}        & \qw                & \multigate{1}{R_2} & \qw \\
& \qw                & \multigate{1}{R_3} & \ghost{R_2}        & \qw \\
& \qw                & \ghost{R_3}        & \qw                & \qw
}}

\caption{\label{reversible} An example of a reversible circuit
  constructed from a single gate $R$. As a product of matrices, we
  write this $R_2 R_3 R_1$, in keeping with the convention
  \cite{Nielsen_Chuang} that circuit diagrams are to be read
  left-to-right, whereas the matrix product acts right-to-left. Note
  that in subsequent circuit diagrams we drop the subscripts from the
  $R$ gates as these can be read off from the ``wires'' the gates act
  on.}
\end{figure}
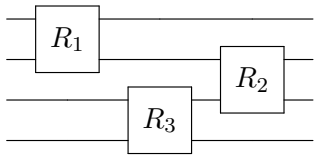

Suppose we arrange $n$ dits on a line, and allow $R$ to act only on
neighboring dits. Further, we do not allow $R$ to be applied
``upside-down''. Then, there are $n-1$ choices for how to apply
$R$. We label these $R_1, R_2, \ldots, R_{n-1}$, as illustrated in
Figure \ref{reversible}. Each of $R_1,\ldots,R_{n-1}$ corresponds to a
$d^n \times d^n$ permutation matrix. Specifically, $R_j$ is obtained
by taking the tensor product of $R$ with identity matrices according to
$R_j = \id_{d \times  d}^{\otimes (j-1)} \otimes R \otimes \id_{d \times d}^{\otimes (n-j-1)}$.

$R_1,\ldots,R_{n-1}$ generate a subgroup of $S_{d^n}$. Among
others, these generators obey the relations
\begin{equation}
R_i R_j = R_j R_i \quad \forall |i-j| \geq 2.
\end{equation}
If $R$ satisfies
\begin{equation}
\label{Yang-Baxter}
R_1 R_2 R_1 = R_2 R_1 R_2
\end{equation}
then
\begin{equation}
R_i R_{i+1} R_i = R_{i+1} R_i R_{i+1} \quad \forall i
\end{equation}
and in this case the gates $R_1,\ldots,R_{n-1}$ satisfy all the
relations of the braid group $B_n$. In other words, the map defined by
$\sigma_i \mapsto R_i$ and $\sigma_i^{-1} \mapsto R_i^{-1}$ is a
homomorphism from $B_n$ to $S_{d^n}$, \emph{i.e.} a representation of
the braid group. Note that this representation is never faithful as
$B_n$ is infinite.

The condition \ref{Yang-Baxter} is known as the Yang-Baxter
equation\footnote{Actually, two slightly different equations go by the
  name Yang-Baxter in the literature. Careful sources
  distinguish these as the algebraic Yang-Baxter equation and the
  braided Yang-Baxter relation (which is sometimes called the quantum
  Yang-Baxter equation). Equation \ref{Yang-Baxter} is the
  latter. Furthermore, some sources treat a more complicated version
  of the Yang-Baxter equation in which $R$ depends on a continuous
  parameter. In such works equation \ref{Yang-Baxter} is often
  referred to as the constant Yang-Baxter equation.}. Finding all the
matrices satisfying the Yang-Baxter equation at a given dimension has
only been achieved at $d=2$ \cite{Hietarinta}. However, certain
systematic constructions coming from mathematical physics can produce
infinite families of solutions. In particular, let $G$ be any finite
group, and let $R$ be the permutation on the set $G \times G$ defined
by
\begin{equation}
%\label{qdouble}
R(a,b) = (b, b^{-1}ab).
\end{equation}
By direct calculation one sees that any such an $R$ satisfies the
Yang-Baxter equation. (In physics language, $R$ comes from the
braiding statistics of the magnetic fluxes in the quantum double of
$G$.)

In 1997, Kitaev discovered that choosing $G$ to be the symmetric group
$S_5$ yields an $R$ gate sufficient to perform universal reversible
computation~\cite{Kitaev03}. Ogburn and Preskill subsequently showed
that the alternating group $A_5$, which is half as large as $S_5$, is
already sufficient. The universality construction for $A_5$ was
subsequently presented in greater detail and generalized to all
non-solvable groups by Mochon~\cite{Mochon}. In the remainder of this
section we give a self-contained exposition of the universality
construction from~\cite{Mochon}, shorn of physics language.

To obtain a representation of the braid group, we must strictly
enforce the requirement that application of $R$ to neighboring dits on
a line is the only allowed operation. In particular, we are not given
as elementary operations the ability to apply $R$ upside-down, or to
non-neighboring dits, or to move dits around. Thus, to prove
computational universality, it is helpful to first construct a
SWAP gate from $R$ gates, which exchanges neighboring dits. As is
well-known, the $n-1$ swaps of nearest neighbors on a line generate
the full group $S_n$ of permutations, and thus a SWAP gate enables
application of $R$ to any pair of dits.

For $R$ gates of the form \eqref{qdouble}, two pairs of inverse group
elements in the order $a,a^{-1},b,b^{-1}$ can be swapped by applying
the product $R_2 R_3 R_1 R_2$. Thus, in the construction
of~\cite{OP99, Mochon}, elements of $A_5$ are always paired 
with their inverses. This can be regarded as a form of encoding;
$|A_5| = 60$, so each 60-state dit is encoded by a corresponding pair
of elements of $A_5$. We introduce the notation $\widetilde{g} \equiv
(g,g^{-1})$ for this encoding, and similarly, abbreviate the encoded
swap operation as follows.
\[
\begin{array}{lcccr}
\begin{array}{c}
\Qcircuit @C=1em @R=.5em {
\lstick{\widetilde{a}} & \multigate{1}{S} & \rstick{\widetilde{b}} \qw \\
\lstick{\widetilde{b}} & \ghost{S}        & \rstick{\widetilde{a}} \qw
}
\end{array} 
& \ & \equiv & \quad &
\begin{array}{c}
\Qcircuit @C=1em @R=.5em {
\lstick{a^{\phantom{-1}}} & \qw              & \multigate{1}{R}   & \qw                & \qw              & \rstick{b} \qw\\
\lstick{a^{-1}}        & \multigate{1}{R} & \ghost{R}          & \qw                & \multigate{1}{R} & \rstick{b^{-1}} \qw \\
\lstick{b^{\phantom{-1}}} & \ghost{R}        & \qw                & \multigate{1}{R}   & \ghost{R}        & \rstick{a} \qw\\
\lstick{b^{-1}}        & \qw              & \qw                & \ghost{R}          & \qw              & \rstick{a^{-1}} \qw
}
\end{array}
\end{array}
\]
Similarly, the sequence $R_2 R_3 R_3 R_2$ performs the
transformation $(\widetilde{a},\widetilde{b}) \mapsto
(\widetilde{a},\widetilde{aba^{-1}})$ on a pair of encoded dits. We
abbreviate this in circuit diagrams as follows.
\[
\begin{array}{lcccl}
\begin{array}{c}
\Qcircuit @C=1em @R=.5em {
\lstick{\widetilde{a}} & \ctrl{1} & \rstick{\widetilde{a}}       \qw \\
\lstick{\widetilde{b}} & \gate{C} & \rstick{\widetilde{aba^{-1}}} \qw 
}
\end{array}
& \quad \quad & \equiv & \quad &
\begin{array}{c} 
\Qcircuit @C=1em @R=.5em {
\lstick{a^{\phantom{-1}}} & \qw              & \qw              & \qw                & \qw              & \rstick{a} \qw\\
\lstick{a^{-1}}        & \multigate{1}{R} & \qw              & \qw                & \multigate{1}{R} & \rstick{a^{-1}} \qw \\
\lstick{b^{\phantom{-1}}} & \ghost{R}        & \multigate{1}{R} & \multigate{1}{R}   & \ghost{R}        & \rstick{aba^{-1}} \qw\\
\lstick{b^{-1}}        & \qw              & \ghost{R}        & \ghost{R}          & \qw              & \rstick{ab^{-1}a^{-1}} \qw
}
\end{array} \\ \\
\begin{array}{c}
\Qcircuit @C=1em @R=.5em {
\lstick{\widetilde{a}} & \ctrl{1} & \rstick{\widetilde{a}}       \qw \\
\lstick{\widetilde{b}} & \gate{C^{-1}} & \rstick{\widetilde{a^{-1}ba}} \qw 
}
\end{array}
& \quad \quad & \equiv & \quad &
\begin{array}{c}
\Qcircuit @C=1em @R=.5em {
\lstick{a^{\phantom{-1}}} & \qw                  & \qw                 & \qw                  & \qw                  & \rstick{a} \qw      \\
\lstick{a^{-1}}        & \multigate{1}{R^{-1}} & \qw                 & \qw                  & \multigate{1}{R^{-1}} & \rstick{a^{-1}} \qw   \\
\lstick{b^{\phantom{-1}}} & \ghost{R^{-1}}        & \multigate{1}{R^{-1}} & \multigate{1}{R^{-1}} & \ghost{R^{-1}}        & \rstick{a^{-1}ba} \qw \\
\lstick{b^{-1}}        & \qw                 & \ghost{R^{-1}}        & \ghost{R^{-1}}        & \qw                  & \rstick{a^{-1}b^{-1}a} \qw
}
\end{array}
\end{array}
\]
This notation can easily be extended to provide a shorthand for the
sequence of gates needed to implement a $C$ gate between
non-neighboring pairs of bits, as illustrated by the following
examples.
\begin{eqnarray*}
\begin{array}{c}
\Qcircuit @C=1em @R=1em {
& \ctrl{3} & \qw \\
& \qw      & \qw \\
& \qw      & \qw \\
& \gate{C} & \qw
}
\end{array}
& \equiv &
\begin{array}{c}
\Qcircuit @C=1em @R=.5em {
& \qw              & \qw              & \ctrl{1} & \qw              & \qw              & \qw \\
& \qw              & \multigate{1}{S} & \gate{C} & \multigate{1}{S} & \qw              & \qw \\
& \multigate{1}{S} & \ghost{S}        & \qw      & \ghost{S}        & \multigate{1}{S} & \qw \\
& \ghost{S}        & \qw              & \qw      & \qw              & \ghost{S}        & \qw
}
\end{array} \\ \\ \\
\begin{array}{c}
\Qcircuit @C=1em @R=.5em {
& \gate{C}  & \qw \\
& \ctrl{-1} & \qw
}
\end{array}
& \equiv &
\begin{array}{c}
\Qcircuit @C=1em @R=.5em {
& \multigate{1}{S} & \ctrl{1}  & \multigate{1}{S} & \qw \\
& \ghost{S}        & \gate{C}  & \ghost{S}        & \qw
}
\end{array}
\end{eqnarray*}

\noindent Next, consider the following product of elements of $A_5$ (which
should be read right-to-left).
\begin{equation}
f(g_1,g_2) = (521)g_1(14352)g_2(124)g_1^{-1}(15342)g_2^{-1}(521)
\end{equation}
One sees that
\begin{eqnarray*}
f((345),(345)) & = & \id \\
f((345),(435)) & = & \id \\
f((435),(345)) & = & \id \\
f((435),(435)) & = & (12)(34)
\end{eqnarray*}
where $\id$ denotes the identity permutation. Furthermore, conjugating
$(345)$ by $(12)(34)$ yields $(435)$, and conversely, conjugating
$(435)$ by $(12)(34)$ yields $(345)$. Thus, we may think of $(345)$ as
an encoded zero and $(435)$ as an encoded one, and we see that
\begin{equation}
f(g_1,g_2) g_0 f(g_1,g_2)^{-1}
\end{equation}
toggles $g_0$ between one and zero if $g_1$ and $g_2$ are both encoded
ones and leaves $g_0$ unchanged otherwise. Such a doubly-controlled
toggling operation is known as a Toffoli gate, which is well-known to
be a computationally universal reversible gate~\cite{Fredkin_Toffoli}.

As a circuit diagram, this construction can be expressed as follows.

\begin{center}
\begin{tabular}{c}
\Qcircuit @C=1em @R=1.5em {
\lstick{\widetilde{(14352)}} & \qw      & \qw         & \qw      & \qw         & \qw      & \qw      & \ctrl{4} & \qw      & \qw     & \rstick{\widetilde{(14352)}} \qw \\
\lstick{\widetilde{(15342)}} & \qw      & \qw         & \ctrl{3} & \qw         & \qw      & \qw      & \qw      & \qw      & \qw     & \rstick{\widetilde{(15342)}} \qw \\
\lstick{\widetilde{(124)}}   & \qw      & \qw         & \qw      & \qw         & \ctrl{2} & \qw      & \qw      & \qw      & \qw     & \rstick{\widetilde{(124)}} \qw \\
\lstick{\widetilde{(521)}}   & \ctrl{1} & \qw         & \qw      & \qw         & \qw      & \qw      & \qw      & \qw      & \ctrl{1}& \rstick{\widetilde{(521)}} \qw \\
\lstick{\widetilde{g_0}}                 & \gate{C} & \gate{C^{-1}} & \gate{C} & \gate{C^{-1}}& \gate{C} & \gate{C} & \gate{C} & \gate{C} & \gate{C}& \rstick{\widetilde{g_0'}} \qw \\
\lstick{\widetilde{g_1}}                 & \qw      & \qw         & \qw       & \ctrl{-1}  & \qw      & \qw      & \qw      & \ctrl{-1}& \qw     & \rstick{\widetilde{g_1}} \qw \\
\lstick{\widetilde{g_2}}                 & \qw      & \ctrl{-2}   & \qw       & \qw        & \qw      & \ctrl{-2}& \qw      & \qw      & \qw     & \rstick{\widetilde{g_2}} \qw
}\end{tabular}\end{center}

\noindent Here, if $g_0,g_1,g_2$ encode bits $b_0,b_1,b_2$ then $g_0'$ encodes
$b_0 \oplus b_1 \land b_2$. The four ancillary dits
$\widetilde{(14352)}$, $\widetilde{(15342)}$, $\widetilde{(124)}$, and
$\widetilde{(521)}$, are used to ``catalytically'' facilitate the
construction of a Toffoli gate, and thus computations built from
arbitrarily many Toffoli gates can be performed with only one copy of
these four dits.

Unpacking the various shorthand notations, one sees that the above
circuit represents the following braid of 132 crossings on 14 strands,
which encodes a Toffoli gate with the first wire as target, and the
second and third wires as controls.
\begin{equation}
\label{explicit}
\begin{array}{ccllll}
T & = &
\sigma_{8} \sigma_{9} \sigma_{9} \sigma_{8}
& \sigma_{10} \sigma_{11} \sigma_{9} \sigma_{10}
& \sigma_{10} \sigma_{11} \sigma_{11} \sigma_{10}
& \sigma_{10} \sigma_{11} \sigma_{9} \sigma_{10} \\
& & \sigma_{2} \sigma_{3} \sigma_{1} \sigma_{2}
& \sigma_{4} \sigma_{5} \sigma_{3} \sigma_{4}
& \sigma_{6} \sigma_{7} \sigma_{5} \sigma_{6}
& \sigma_{8} \sigma_{9} \sigma_{9} \sigma_{8} \\
& & \sigma_{6} \sigma_{7} \sigma_{5} \sigma_{6}
& \sigma_{4} \sigma_{5} \sigma_{3} \sigma_{4}
& \sigma_{2} \sigma_{3} \sigma_{1} \sigma_{2}
& \sigma_{12} \sigma_{13} \sigma_{11} \sigma_{12}\\
& & \sigma_{10} \sigma_{11} \sigma_{9} \sigma_{10}
& \sigma_{10} \sigma_{11} \sigma_{11} \sigma_{10}
& \sigma_{10} \sigma_{11} \sigma_{9} \sigma_{10}
& \sigma_{12} \sigma_{13} \sigma_{11} \sigma_{12}\\
& & \sigma_{6} \sigma_{7} \sigma_{5} \sigma_{6}
& \sigma_{8} \sigma_{9} \sigma_{9} \sigma_{8}
& \sigma_{6} \sigma_{7} \sigma_{5} \sigma_{6}
& \sigma_{10} \sigma_{11} \sigma_{9} \sigma_{10}\\
& & \sigma^{-1}_{10} \sigma^{-1}_{11} \sigma^{-1}_{11} \sigma^{-1}_{10}
& \sigma_{10} \sigma_{11} \sigma_{9} \sigma_{10}
& \sigma_{4} \sigma_{5} \sigma_{3} \sigma_{4}
& \sigma_{6} \sigma_{7} \sigma_{5} \sigma_{6}\\
& & \sigma_{8} \sigma_{9} \sigma_{9} \sigma_{8}
& \sigma_{6} \sigma_{7} \sigma_{5} \sigma_{6}
& \sigma_{4} \sigma_{5} \sigma_{3} \sigma_{4}
& \sigma_{12} \sigma_{13} \sigma_{11} \sigma_{12}\\
& & \sigma_{10} \sigma_{11} \sigma_{9} \sigma_{10}
& \sigma^{-1}_{10} \sigma^{-1}_{11} \sigma^{-1}_{11} \sigma^{-1}_{10}
& \sigma_{10} \sigma_{11} \sigma_{9} \sigma_{10}
& \sigma_{12} \sigma_{13} \sigma_{11} \sigma_{12}\\
& & \sigma_{8} \sigma_{9} \sigma_{9} \sigma_{8}
\end{array}
\end{equation}
Note that we take the convention that this should be read backwards
compared to the way one reads English text. This is in keeping with
the conventional notation for the composition of functions and our
right-to-left multiplication of $R$ matrices. We have used whitespace
to divide crossings into groups of four as these correspond to
elementary $S$ and $R$ gates.

Given this construction of the Toffoli gate by braid crossings, it is
a simple matter to ``compile'' any given logic circuit into a
corresponding braid. $A_5$ has 60 elements. Thus, encoding a single
bit into a a pair of $A_5$ elements appears somewhat wasteful. It is
natural to try to find Yang-Baxter solutions acting on $d$-state dits
for smaller $d$ that achieve universal classical computation. In
appendix \ref{optimizing}, we improve upon the $A_5$-based
construction to show that $d=44$ suffices. We have also used
exhaustive computer search to find all permutation solutions
satisfying the Yang-Baxter equation up to $d=5$ (i.e. up to $25 \times
25$ permutation matrices). Our examination of these solutions suggests
that none are computationally universal. Where between $5$ and $44$
lies the minimal $d$ remains an interesting open question.

\section{Optimizing Classical Braid Gates}
\label{optimizing}

In appendix \ref{universal} we have recounted the construction of
\cite{Mochon}, which shows that the reversible gate $R$, which acts on
pairs of 60-state dits and satisfies the Yang-Baxter equation, can
perform universal classical computation. In this section, based on a
suggestion of Robert K\"onig, we show that $R$ can be modified to
obtain a gate acting on pairs of 44-state dits that satisfies the
Yang-Baxter equation and can perform universal classical
computation. Our computational evidence suggests that no reversible
gate on $d$-state dits satisfying the Yang-Baxter equation can perform
universal computation for $d \leq 5$. Where between 5 and 44 the
minimal $d$ lies for which computationally universal reversible
Yang-Baxter gates acting on $d$-state qudits exist remains an open
question.

The universality construction of \cite{Mochon}, recounted in appendix
\ref{universal}, starts with all dits initialized to states from the
following set.
\begin{eqnarray*}
S & = & \{g,g^{-1}|g \in S_0\} \\
S_0 & = & \{(14352), (15342), (124), (521), (345), (435)\}
\end{eqnarray*}
Here we show that the orbit of $S$ under the action of the gate $R$ is
not all of $A_5$, rather the orbit has only 44 elements. Thus the
restriction of the matrix $R$ onto this 44-dimensional subspace is a
permutation-matrix that satisfies the Yang-Baxter equation and is
capable of universal classical computation.

Recalling \eqref{qdouble}, one sees that the orbit $O_R$ of $S$ under
$R$ is 
\begin{equation}
O_R = \{b^{-1}ab|a \in S, b \in \langle S \rangle \}
\end{equation}
where $\langle S \rangle$ is the subgroup of $A_5$ generated by $S$. A
simple computer algebra calculation shows that $\langle S \rangle =
A_5$, thus $O_R$ consists of exactly those elements of $A_5$ conjugate
to $S$.

It is well known that the conjugacy classes of $A_5$ are as
follows.
\begin{center}
\begin{tabular}{ll}
1) the identity & (1 element) \\
2) 3-cycles & (20 elements) \\
3) conjugates of (12)(34) & (15 elements) \\
4) conjugates of (12345) & (12 elements) \\
5) conjugates of (21345) & (12 elements)
\end{tabular}
\end{center}
One sees that $O_R$ contains 2), and does not contain 1) or 3). The
only remaining question is whether $O_R$ contains both 4) and 5) or
just one of them. A simple computer algebra calculation shows that
(14352) and (15342) are non-conjugate elements of $A_5$. Hence $O_R$
must contain both 4) and 5). Therefore, $|O_R| = 44$. 

\bibliography{obf}

\begin{thebibliography}{10}

\bibitem{Aaronson}
Scott Aaronson.
\newblock {BQP} and the polynomial hierarchy.
\newblock In {\em STOC '10: Proceedings of the 42nd ACM symposium on Theory of
  Computing}, pages 141--150, 2010.
\newblock {\tt arXiv:0910.4698}.

\bibitem{AA11}
Dorit Aharonov and Itai Arad.
\newblock The {BQP}-hardness of approximating the {J}ones polynomial.
\newblock {\em New Journal of Physics}, 13(3):035019, 2011.

\bibitem{ABE08}
Dorit Aharonov, Michael Ben-Or, and Elad Eban.
\newblock Interactive proofs for quantum computation.
\newblock In {\em Proceedings of Innovations in Computer Science (ICS 2010)},
  pages 453--469, 2010.
\newblock arXiv:0810.5375.

\bibitem{ABJ12}
G.~Alagic, S.~Jordan, and A.~Bapat.
\newblock Classical simulation of {Y}ang-{B}axter gates.
\newblock To appear in: Proceedings of TQC2014.

\bibitem{Alagic}
Gorjan Alagic, Stephen~P. Jordan, Robert Koenig, and Ben~W. Reichardt.
\newblock Approximating {T}uraev-{V}iro 3-manifold invariants is universal for
  quantum computation.
\newblock {\em Physical Review A}, 82:040302(R), 2010.
\newblock {\tt arXiv:1003.0923}.

\bibitem{Artin25}
Emil Artin.
\newblock Theorie der z{\"o}pfe.
\newblock {\em Abhandlungen aus dem Mathematischen Seminar der Universit{\"a}t
  Hamburg}, 4:42--72, 1925.

\bibitem{Barak}
B.~Barak.
\newblock Can we obfuscate programs?
\newblock \url{http://www.cs.princeton.edu/~boaz/Papers/obf_informal.html}.

\bibitem{BGIRSVY01}
Boaz Barak, Oded Goldreich, Russell Impagliazzo, Steven Rudich, Amit Sahai,
  Salil~P. Vadhan, and Ke~Yang.
\newblock On the (im)possibility of obfuscating programs.
\newblock In {\em Advances in Cryptology - CRYPTO 2001}, number 2139 in Lecture
  Notes in Computer Science, pages 1--18. Springer-Verlag, 2001.

\bibitem{Bennett}
C.~H. Bennett.
\newblock Logical reversibility of computation.
\newblock {\em IBM Journal of Research and Development}, 17(6):525--532, 1973.

\bibitem{BFK08}
Anne Broadbent, Joseph Fitzsimons, and Elham Kashefi.
\newblock Universal blind quantum computation.
\newblock In {\em Proceedings of the 50th Annual IEEE Symposium on Fountations
  of Computer Science (FOCS 2008)}, pages 517--526, 2009.
\newblock arXiv:0807.4154.

\bibitem{CT02}
Christian~S. Collberg and Clark Thomborson.
\newblock Watermarking, tamper-proofing, and obfuscation -- tools for software
  protection.
\newblock {\em IEEE Transactions on Software Engineering}, 28(8):735--746,
  2002.

\bibitem{Coxeter}
H.~S.~M. Coxeter.
\newblock Factor groups of the braid group.
\newblock In {\em Proceedings of the 4th Canadian Mathematical Congress}, pages
  95--122, 1959.
\newblock See \texttt{http://mathoverflow.net/questions/48849/}.

\bibitem{Dawson_Nielsen}
Christopher~M. Dawson and Michael~A. Nielsen.
\newblock The {S}olovay-{K}itaev algorithm.
\newblock {\em Quantum Information and Computation}, 6(1):81--95, 2006.
\newblock arXiv:quant-ph/0505030.

\bibitem{Dehornoy08}
Patrick Dehornoy.
\newblock Efficient solutions to the braid isotopy problem.
\newblock {\em Discrete Applied Mathematics}, 156:3094--3112, 2008.
\newblock {\tt arxiv:math/0703666}.

\bibitem{Epstein92}
D.~Epstein, J.~Cannon, D.~Holt, S.~Levy, M.~Paterson, and W.~Thurston.
\newblock {\em Word processing in groups}.
\newblock Jones and Bartlett Publ., 1992.

\bibitem{Fefferman}
Bill Fefferman and Chris Umans.
\newblock Pseudorandom generators and the {BQP} vs. {PH} problem, 2010.
\newblock {\tt arXiv:1007.0305}.

\bibitem{Fredkin_Toffoli}
E.~Fredkin and T.~Toffoli.
\newblock Conservative logic.
\newblock {\em International Journal of Theoretical Physics}, 21(3/4):219--253,
  1982.

\bibitem{FLW02}
Michael~H. Freedman, Michael Larsen, and Zhenghan Wang.
\newblock A modular functor which is universal for quantum computation.
\newblock {\em Communications in Mathematical Physics}, 227:605--622, 2002.
\newblock {\tt arXiv:quant-ph/0001108}.

\bibitem{GGHRSW13}
Sanjam Garg, Craig Gentry, Shai Halevi, Mariana Raykova, Amit Sahai, and Brent
  Waters.
\newblock Candidate indistinguishability obfuscation and functional encryption
  for all circuits.
\newblock In {\em Proceedings of the 54th Annual IEEE Symposium on Foundations
  of Computer Science (FOCS)}, pages 40--49, 2013.

\bibitem{Garside69}
F.A. Garside.
\newblock The braid group and other groups.
\newblock {\em Quart. J. Math. Oxford Ser.}, 2, 20:235--254, 1969.

\bibitem{GR07}
Shafi Goldwasser and Guy~N. Rothblum.
\newblock On best-possible obfuscation.
\newblock In {\em Theory of Cryptography - TCC 2007}, pages 194--213. Springer,
  2007.

\bibitem{Gonzalez10}
Juan Gonz{\'a}lez-Meneses.
\newblock Basic results on braid groups, 2010.
\newblock {\tt arxiv:1010.0321 [math]}.

\bibitem{Hamidi-Tehrani}
Hessam Hamidi-Tehrani.
\newblock On complexity of the word problem in braid groups and mapping class
  groups.
\newblock {\em Topology and its Applications}, 105:237--259, 2000.

\bibitem{Hietarinta}
Jarmo Hietarinta.
\newblock All solutions to the constant quantum {Y}ang-{B}axter equation in two
  dimensions.
\newblock {\em Physics Letters A}, 165:245--251, 1992.

\bibitem{HS03}
D.~Hofheinz and R.~Steinwandt.
\newblock A practical attack on some braid group based cryptographic
  primitives.
\newblock In {\em Public Key Cryptography}, pages 187--198, 2003.

\bibitem{Huntington}
Edward~V. Huntington.
\newblock Sets of independent postulates for the algebra of logic.
\newblock {\em Transactions of the American Mathematical Society}, 4:288--309,
  1904.

\bibitem{Iwama}
Kazuo Iwama, Yahiko Kambayashi, and Shigeru Yamashita.
\newblock Transformation rules for designing {CNOT}-based quantum circuits.
\newblock In {\em DAC '02: Proceedings of the 39th annual Design Automation
  Conference}, pages 419--424, 2002.

\bibitem{JWB03}
Dominik Janzing, Pawel Wocjan, and Thomas Beth.
\newblock {``Identity Check'' is QMA-complete}, 2003.
\newblock {\tt arXiv:quant-ph/0305050}.

\bibitem{Jordan13}
Stephen Jordan.
\newblock Strong equivalence of reversible circuits is {coNP}-complete.
\newblock {\em Quantum Information and Computation}, 14(15/16):1303--1308,
  2014.
\newblock arXiv:1307.0836.

\bibitem{Kauffman91}
Louis~H. Kauffman.
\newblock {\em Knots and Physics}.
\newblock Wold Scientific, 1991.

\bibitem{Kitaev03}
A.~Yu. Kitaev.
\newblock Fault-tolerant quantum computation by anyons.
\newblock {\em Annals of Physics}, 303:2--30, 2003.
\newblock {\tt arXiv:quant-ph/9707021}.

\bibitem{Krovi_Russell}
Hari Krovi and Alexander Russell.
\newblock Quantum fourier transforms and the complexity of link invariants for
  quantum doubles of finite groups, 2012.
\newblock {\tt arXiv:quant-ph/1210.1550 [quant-ph]}.

\bibitem{Marriott_Watrous}
Chris Marriott and John Watrous.
\newblock Quantum {A}rthur-{M}erlin games.
\newblock {\em Computational Complexity}, 14(2):122--152, 2005.
\newblock arXiv:cs/0506068.

\bibitem{Mochon}
Carlos Mochon.
\newblock Anyons from nonsolvable finite groups are sufficient for universal
  quantum computation.
\newblock {\em Physical Review A}, 67(2):022315, 2003.
\newblock {\tt arXiv:quant-ph/0206128}.

\bibitem{Nagaj_Wocjan_Zhang}
Daniel Nagaj, Pawel Wocjan, and Yong Zhang.
\newblock Fast amplification of {QMA}.
\newblock {\em Quantum Information and Computation}, 9(11/12):1053--1068, 2009.
\newblock arXiv:0904.1549.

\bibitem{Nielsen_Chuang}
Michael~A. Nielsen and Isaac~L. Chuang.
\newblock {\em Quantum computation and quantum information}.
\newblock Cambridge University Press, 2000.

\bibitem{OP99}
R.~Walter Ogburn and John Preskill.
\newblock Topological quantum computation.
\newblock In {\em Quantum Computing and Quantum Communications}, volume 1509 of
  {\em Lecture Notes in Computer Science}, pages 341--356. Springer, 1999.
\newblock First NASA International Conference QCQC '98.

\bibitem{SahaiW13}
Amit Sahai and Brent Waters.
\newblock How to use indistinguishability obfuscation: Deniable encryption, and
  more.
\newblock {\em IACR Cryptology ePrint Archive}, 2013:454, 2013.

\bibitem{SB09}
Dan Shepherd and Michael~J. Bremner.
\newblock Temporally unstructured quantum computation.
\newblock {\em Proceedings of the Royal Society A}, 465:1413--1439, 2009.
\newblock arXiv:0809.0847.

\bibitem{Shor_Jordan}
Peter~W. Shor and Stephen~P. Jordan.
\newblock Estimating {J}ones polynomials is complete for one clean qubit.
\newblock {\em Quantum Information and Computation}, 8(8/9):681--714, 2008.
\newblock {\tt arXiv:0707.2831}.

\bibitem{Simonaire}
Eric~D. Simonaire.
\newblock Sub-circuit selection and replacement algorithms modeled as term
  rewriting systems.
\newblock Master's thesis, Air Force Institute of Technology, 2008.

\bibitem{Trebst}
Simon Trebst, Matthias Troyer, Zhenghan Wang, and Andreas W.~W. Ludwig.
\newblock A short introduction to {F}ibonacci anyon models.
\newblock {\em Progress in Theoretical Physics Supplement}, 176:384--407, 2008.
\newblock {\tt arXiv:0902.3275}.

\end{thebibliography}

\end{document}